\documentclass[10pt,conference]{IEEEtran}
\IEEEoverridecommandlockouts
\usepackage{authblk}
\usepackage{cite}
\usepackage{amsmath,amssymb,amsfonts}
\usepackage{algorithmic}
\usepackage{graphicx}
\usepackage{textcomp}
\usepackage{xcolor}
\usepackage{multirow}
\usepackage{booktabs}
\usepackage{array}
\usepackage{url}
\usepackage[hidelinks]{hyperref}
\def\BibTeX{{\rm B\kern-.05em{\sc i\kern-.025em b}\kern-.08em
    T\kern-.1667em\lower.7ex\hbox{E}\kern-.125emX}}
    
\begin{document}

\title{Automating Quantum Software Maintenance: Flakiness Detection and Root Cause Analysis}

\author[1]{Janakan Sivaloganathan*} 
\author[2]{Ainaz Jamshidi*\thanks{*The first two coauthors have made equal contributions to this work.}} 
\author[1]{Andriy Miranskyy}
\author[2]{Lei Zhang}

\affil[1]{Department of Computer Science, Toronto Metropolitan University, Toronto, Canada}
\affil[2]{Department of Information Systems, University of Maryland, Baltimore County, USA}

\affil[ ]{jsiva@torontomu.ca, ainazj1@umbc.edu, avm@torontomu.ca, leizhang@umbc.edu} 

\maketitle

\begin{abstract}
Flaky tests, which pass or fail inconsistently without code changes, are a major challenge in software engineering in general and in quantum software engineering in particular due to their complexity and probabilistic nature, leading to hidden issues and wasted developer effort.

We aim to create an automated framework to detect flaky tests in quantum software and an extended dataset of quantum flaky tests, overcoming the limitations of manual methods.

Building on prior manual analysis of 14 quantum software repositories, we expanded the dataset and automated flaky test detection using transformers and cosine similarity. We conducted experiments with Large Language Models (LLMs) from the OpenAI GPT and Meta LLaMA families to assess their ability to detect and classify flaky tests from code and issue descriptions.

Embedding transformers proved effective: we identified 25 new flaky tests, expanding the dataset by 54\%. Top LLMs achieved an F1-score of 0.8871 for flakiness detection but only 0.5839 for root cause identification.

We introduced an automated flaky test detection framework using machine learning, showing promising results but highlighting the need for improved root cause detection and classification in large quantum codebases. Future work will focus on improving detection techniques and developing automatic flaky test fixes.
\end{abstract}

\section{Introduction}

Flaky tests, which exhibit non-deterministic behavior by failing or passing inconsistently without any changes to the code under test, pose significant challenges for software maintenance and reliability~\cite{luo2014empirical}. In the field of quantum software engineering, flaky tests are particularly problematic due to the inherent complexities and probabilistic nature of quantum computations. They can obscure genuine issues, waste developers' time, and undermine confidence in test suites~\cite{zhang2023identifying,zhang2024automated}.

In previous work, Zhang et al.~\cite{zhang2023identifying} explore the code and bug-tracking repositories of 14 quantum software and identify 46 unique quantum flaky tests in 12 quantum projects (ranging from 0.26\% to 1.85\% of bug reports). 
They search for the 10 keywords related to flaky tests (e.g., flaky and flakiness) in issue reports (IRs) and pull requests (PRs) to identify flaky tests. They then identify and categorize eight types of flakiness and seven common fixes. Randomness is the most common cause of quantum flakiness, and the most common solution is to fix pseudo-random number generator (PRNG) seeds. However, the findings (46 instances of flakiness) are constrained by the limitations inherent in their vocabulary-based method. In addition, they manually examine and identify all flaky tests, which is time-consuming. 

Our research explores a more effective and efficient flaky test detection technique for quantum software by answering the following research questions. 
\begin{itemize}
    \item \textbf{RQ1}: How can we detect if a given IR or PR is related to a flaky test?
    \item \textbf{RQ2}: How can we detect if a given IR or PR is related to a flaky test with additional code context?
    \item \textbf{RQ3}: How can we identify the root cause of a flaky test?
\end{itemize}
Our main \textbf{contributions} are as follows.
\begin{itemize}
    \item We have enriched the existing dataset of flaky tests~\cite{zhang2023identifying} by adding more observations, as well as including the buggy code causing flakiness and the corresponding fixes, which were absent from the original dataset. The extended dataset is publicly available at: \url{https://doi.org/10.5281/zenodo.13913775}.
    \item We developed a method to semi-automatically detect flaky test-related IRs and PRs by mining software repositories using embedding transformers and cosine similarity.
    \item We propose a method to automatically detect flaky issues using LLMs with zero-shot prompting.
\end{itemize}

By automating the detection of flaky tests, our framework aims to improve the reliability and maintainability of quantum software systems.

\section{Method}

\subsection{Dataset}

Our study builds upon a prior manual analysis by Zhang et al.~\cite{zhang2023identifying}, who examined 14 open-source quantum software repositories from platforms such as IBM Qiskit, Microsoft Quantum Development Kit, TensorFlow Quantum, and NetKet. They identified 46 flaky test reports across 12 repositories by searching closed GitHub issues using keywords related to flakiness (e.g., flaky, intermittent, nondeterministic).

\subsubsection{Using transformers and cosine similarity for flaky tests detection}

To extend this research and detect more flaky reports systematically and automatically, we employed embedding transformers to represent GitHub IRs and PRs and measured the cosine similarity with previously identified flaky tests.

Based on the Hugging Face leaderboard~\cite{muennighoff2022mteb}, we selected three top-performing (at the time of experiment design) embeddings on generic tasks. We utilized the pre-trained `mixedbread-ai/mxbai-embed-large-v1' transformer~\cite{emb2024mxbai,li2023angle} from Hugging Face library~\cite{wolf2019huggingface} to extract contextual embeddings\footnote{The embeddings were derived from the penultimate layer of the model, ensuring that they captured the nuanced features of the text.} of the GitHub IRs and PRs. We also evaluated other transformers, such as `SFR-Mistral'~\cite{SFRAIResearch2024} and `e5-mistral-7b-instruct transformers'~\cite{wang2023improving,wang2022text}, using $k$-means clustering~\cite{macqueen1967some} to assess their effectiveness in distinguishing between flaky and non-flaky test cases. Our analysis showed that the ``mixedbread-ai/mxbai-embed-large-v1'' model provided the most distinct and separable representation for this task.

Using this model, we generated embeddings from the tokenized text of all scraped GitHub IRs and PRs from the 12 repositories, as well as from the previously identified flaky cases. We calculated cosine similarity scores between the embeddings and ranked the issues based on their similarity to known flaky cases. Excluding identical matches from our initial flaky set, at least two authors cross-examined the top-ranked issues, associated PRs, and code commits to determine if they were related to flaky tests and to establish their cause categories.

In the first iteration, we identified 15 new flaky tests from the top-ranked cosine similarity scores. We then augmented the original dataset with these new cases and repeated the process, resulting in the detection of an additional 10 flaky tests in the second iteration. In total,\textit{ we identified 25 new flaky tests across the 12 repositories, increasing the original dataset size by 54\%.}

To create a balanced dataset, we also collected non-flaky tests from GitHub reports during the cosine similarity analysis. Test cases with a cosine similarity score of less than 0.5 were labeled as non-flaky. Additionally, we included instances that our method incorrectly labeled as flaky to further challenge the classification task. In total, we compiled 71 non-flaky cases to match the size of our expanded flaky dataset.

\subsubsection{Dataset preparation}\label{sec:dataset_prep}
We extracted descriptions and comments from IRs and PRs for each issue observation in the dataset, recorded code differences, and noted affected files before and after fixes. Method-level code changes were also extracted for a more concentrated analysis.

Manual verification ensured that the extracted artifacts matched the identified IRs and PRs. The dataset was organized into `full' and `method' directories, each further divided into `flaky' and `non-flaky' sections.

\subsection{Detecting Flaky Tests with LLMs}

\begin{figure*}
    \centering
    \includegraphics[width=1.0\linewidth]{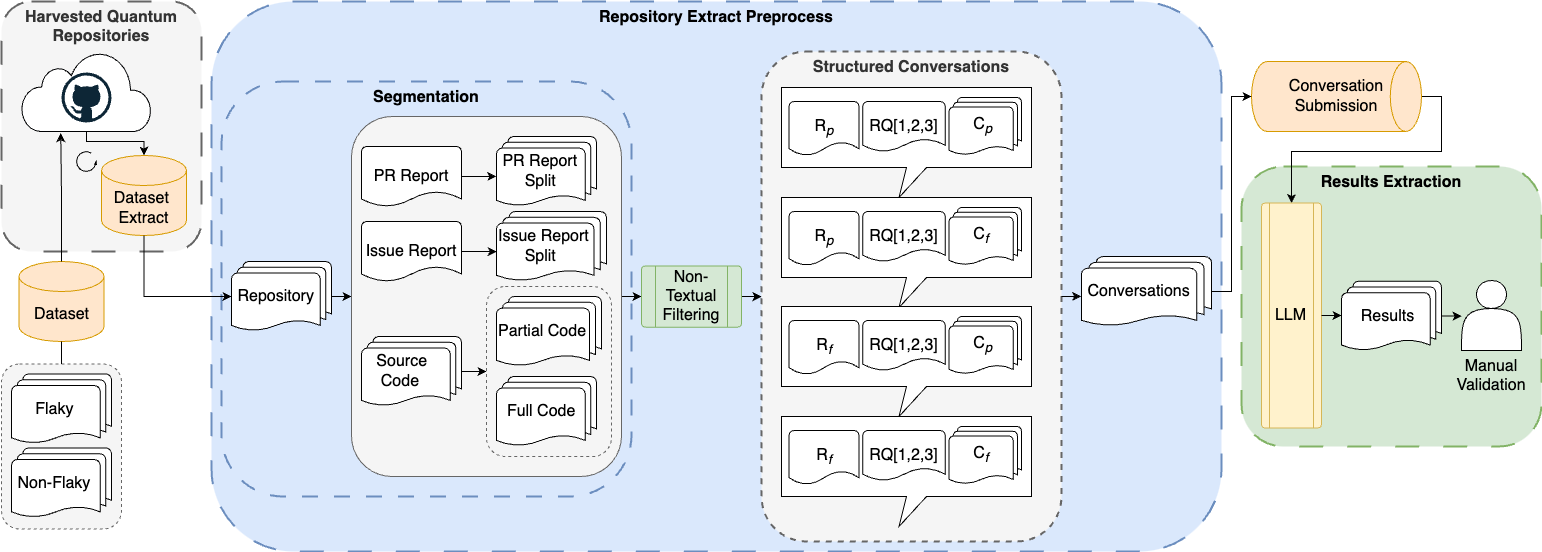}
    \caption{Architecture of the automated framework for quantum flakiness detection.}
    \label{fig:diagram}
\end{figure*}

Our experiments demonstrate an automated LLM-based framework (see Figure~\ref{fig:diagram}) that efficiently gathers resources, configures inputs, and classifies bugs by streamlining the standard software development workflow with GitHub as the version control system. This framework simulates a typical software engineering process for bug resolution.

\subsubsection{LLM Inference Configuration}

Leveraging the extracted codebase data (discussed in Section~\ref{sec:dataset_prep}), we crafted input prompts to address our research questions. The prompts are provided in the supplementary material (\url{https://doi.org/10.5281/zenodo.13913775}).

To explore our research questions and assess how context size affects the answers, we designed the following experiments.

For RQ1, which aims to classify whether a particular IR (or PR, if no issue is associated with it) is flaky or non-flaky, we tested two levels of context: $R_p$ (partial), which includes only the initial IR (or PR) description, and $R_f$ (full), which includes the description along with all associated comments.

For RQ2, we expanded the context for the language model by adding the code involved in the PR before the fix was applied, also at two levels: $C_p$ (partial), which includes the method-level code, and $C_f$ (full), which provides the complete code listing.

By combining the context levels from RQ1 and RQ2, we generated four experimental conditions: $\{R_p, R_f\} \times \{C_p, C_f\}$. These conditions range from $(R_p, C_p)$, which uses only the description and method-level code, to $(R_f, C_f)$, which includes the description with comments and the full code listing.

For RQ3, the amount of information provided did not change; we simply followed up by asking which specific root cause a particular flaky test relates to, using the nine classes of root causes defined by~\cite{zhang2023identifying}: ``Randomness (PRNG),'' ``Floating Point Operations,'' ``Software Environment,'' ``Multi-threading,'' ``Visualization,'' ``Unhandled Exceptions,'' ``Network,'' ``Unordered Collection,'' or ``Others''. We find that we do not need to alter these classes for our extended dataset.

We utilized LangChain~\cite{LangChai97:online}, an integration framework that abstracts prompt templating, retrieval strategies, and chaining, allowing us to manage conversational memory effectively. This setup enabled us to simulate a developer-to-AI interaction, appending additional context to study its impact on the LLM's reasoning.

\subsubsection{LLMs under study} 
To assess the performance of various LLMs, we study two open-source LLMs, namely Meta LLaMA-70B-Instruct v.3.1, and LLaMA-405B-Instruct v.3.1~\cite{dubey2024LLaMA}, and two closed-source LLMs, namely OpenAI GPT-4o-mini v.2024-10-03 and GPT-4o v.2024-10-07~\cite{openai2024gpt4omini,openai2024gpt4o}. All captured OpenAI and Meta Instruct models were accessed remotely through serverless APIs of OpenAI~\cite{Overview51:online} and Google VertexAI~\cite{VertexAI96:online}, respectively.\footnote{We also fine-tuned CodeBERT~\cite{feng2020codebert} for flaky test detection using a few-shot learning approach. While training was successful, the model failed to generalize effectively for the test set when applied to GitHub IRs and PRs.}

\subsection{Limitations}

\subsubsection{Repository Differences}
In cases where the issue and fix resided in different repositories, we manually aligned the IRs with the fix repositories to maintain consistency in our analysis.

\subsubsection{Multiple Pull Requests}
In some cases of the dataset, a flaky bug will have two PRs associated with it. For the four observations in the flaky data and two in the non-flaky data, we split the observation into two two instances. A single issue that is observed in both PRs will be present in both of the codebases respectively.

\subsubsection{Partial Code Extraction (Method-Level)}\label{sec:code_extraction}
The differences between methods were automatically extracted using PyDriller~\cite{Spadini2018}, which operates only on Python code. Given that Python dominates our dataset (12 out of 14 repositories are written primarily in Python), this limitation is not a significant concern.

However, even in Python-focused datasets, not all repositories contain method-level data. For example, some fixes might target configuration files or global variables within Python scripts. Therefore, we capture and report the total number of observations for each experimental setup to account for such cases.

\section{Results and Analysis}

\begin{table*}[htbp]
\centering
\small
\caption{Model Performance Comparison}
\label{tab:results}
\begin{tabular}{@{}llcccccccc@{}}
\toprule
\multirow{2}{*}{Model} & \multirow{2}{*}{Context} & \multicolumn{2}{c}{F1} & \multicolumn{2}{c}{MCC} & \multicolumn{2}{c}{Recall} & \multicolumn{2}{c}{Total Observations} \\
\cmidrule(lr){3-4} \cmidrule(lr){5-6} \cmidrule(lr){7-8} \cmidrule(lr){9-10}
 &  & RQ1 & RQ3 & RQ1 & RQ3 & RQ1 & RQ2 & RQ1 & RQ2\&3 \\
\midrule
\multirow{4}{*}{GPT-4o} & $\{R_p, C_p \}$ & \multirow{2}{*}{0.8443} & \textbf{0.5839} & \multirow{2}{*}{0.6971} & \textbf{0.5928} & \multirow{2}{*}{0.7746} & 0.7955 & 142 & 44 \\
 & $\{R_p, C_f \}$ & & 0.4316 & & 0.3823 & & 0.6290 & 142 & 62 \\
 & $\{R_f, C_p \}$ & \multirow{2}{*}{0.8731} & 0.5562 & \multirow{2}{*}{0.7477} & 0.5689 & \multirow{2}{*}{0.8451} & \textbf{0.8636} & 142 & 44 \\
 & $\{R_f, C_f \}$ & & 0.4919 & & 0.4805 & & 0.7580 & 142 & 62 \\
\midrule
\multirow{4}{*}{LLaMA-405B-Instruct} & $\{R_p, C_p \}$ & \multirow{2}{*}{0.8163} & 0.5219 & \multirow{2}{*}{0.6379} & 0.5457 & \multirow{2}{*}{0.7606} & 0.7727 & 142 & 44 \\
 & $\{R_p, C_f \}$ & & 0.4271 & & 0.4252 & & 0.5806 & 142 & 62 \\
 & $\{R_f, C_p \}$ & \multirow{2}{*}{0.8519} & 0.5251 & \multirow{2}{*}{0.7060} & 0.5515 & \multirow{2}{*}{0.8169} & 0.7727 & 142 & 44 \\
 & $\{R_f, C_f \}$ & & 0.4729 & & 0.4855 & & 0.6935 & 142 & 62 \\
\midrule
\multirow{4}{*}{GPT-4o-mini} & $\{R_p, C_p \}$ & \multirow{2}{*}{0.8229} & 0.5219 & \multirow{2}{*}{0.6558} & 0.5606 & \multirow{2}{*}{0.7465} & 0.6136 & 142 & 44 \\
 & $\{R_p, C_f \}$ & & 0.4729 & & 0.4874 & & 0.5000 & 142 & 62 \\
 & $\{R_f, C_p \}$ & \multirow{2}{*}{\textbf{0.8871}} & 0.5306 & \multirow{2}{*}{\textbf{0.7774}} & 0.5659 & \multirow{2}{*}{\textbf{0.8451}} & 0.6818 & 142 & 44 \\
 & $\{R_f, C_f \}$ & & 0.4754 & & 0.4980 & & 0.6452 & 142 & 62 \\
\midrule
\multirow{4}{*}{LLaMA-70B-Instruct} & $\{R_p, C_p \}$ & \multirow{2}{*}{0.8351} & 0.5158 & \multirow{2}{*}{0.7016} & 0.5230 & \multirow{2}{*}{0.7042} & 0.5227 & 142 & 44 \\
 & $\{R_p, C_f \}$ & & 0.4430 & & 0.4368 & & 0.5806 & 142 & 62 \\
 & $\{R_f, C_p \}$ & \multirow{2}{*}{0.7980} & 0.5333 & \multirow{2}{*}{0.6370} & 0.5505 & \multirow{2}{*}{0.6479} & 0.5682 & 142 & 44 \\
 & $\{R_f, C_f \}$ & & 0.4760 & & 0.4895 & & 0.6452 & 142 & 62 \\
\bottomrule
\multicolumn{10}{@{}l@{}}{\small Note: For RQ1, the $C_{(\cdot)}$ contexts are not used, hence the identical values for both contexts.}
\end{tabular}
\end{table*}

\subsection{LLM Detection: Interpretability and Challenges}   
We adopted four LLMs GPT-4o, GPT-4o-mini, LLaMA-405B-Instruct, and LLaMA-70B-Instruct, and evaluated the performance of the LLM in classifying RQ1 based on both flaky and non-flaky observations, and RQ2 and RQ3 based solely on the ground truths of the flaky observations. The results can be found in Table~\ref{tab:results}; we employ the F1-score, Mathews Correlation Coefficient (MCC), recall, and the number of detected flaky/non-flaky tests to compare the performance.

We also evaluated two additional LLMs on-site using the Ollama framework~\cite{ollama2024}. However, the results for the non-instruct tuned LLaMA-8B and LLaMA-70B models were excluded due to insufficient performance, as many outputs were either empty or corrupted. This underperformance is likely due to these being the smallest non-fine-tuned models in our study. Additionally, GPT-4o-mini was the only model that required explicit manual modifications before scoring.

The difference in the number of observations between the $C_p$ and $C_f$ contexts is attributed to cases where method-level code could not be extracted. These situations include cases where the code was not written in Python or when no changes affected the methods in the code.

\subsection{Model Performance}

\subsubsection{RQ1}

Using the full context, the best-performing model is GPT-4o-mini, with an F1-score of 0.8871 and an MCC of 0.7774. This outcome is somewhat unexpected, as GPT-4o is generally considered more powerful. However, when only partial context is provided, GPT-4o outperforms GPT-4o-mini with an F1-score of 0.8443 vs. 0.8229 and MCC of 0.6971 vs. 0.6558. Thus, GPT-4o effectively identifies flakiness with respectable performance when provided just the initial report. This suggests that classification with limited context might be more challenging, and a more sophisticated model excels in such cases.
LLaMA-405B-Instruct and LLaMA-70B-Instruct ranked third or fourth (depending on the context).

\subsubsection{RQ2}
We assess whether adding code at the method-level ($C_p$) or including a full code listing ($C_f$) improves classification accuracy. We compare recall values from RQ1 and RQ2 to evaluate this. For GPT-4o, adding method-level code ($C_p$) increases recall, with the highest score observed in the $\{R_f, C_p\}$ setup, as expected. Interestingly, providing a complete code listing ($C_f$) reduces recall, which aligns with the observation that a model, much like a human programmer, might struggle to identify specific methods to focus on when faced with the entire codebase.

For GPT-4o-mini, recall decreases for RQ2, but less so for $C_p$ compared to $C_f$. This suggests that analyzing both natural language and code is a more complex task, one that requires a more advanced model. 

The behavior of LLaMA-405B-Instruct is similar to GPT-4o, showing an increase in recall for $C_p$ and a decrease in recall for $C_f$. In contrast, LLaMA-70B-Instruct exhibits behavior similar to GPT-4o-mini, with decreased performance in both contexts.

Note that we should be cautious about drawing strong conclusions when comparing $C_p$ and $C_f$, as the number of observations for the $C_p$ cases is smaller than for the $C_f$ cases (44 vs. 62, respectively, due to the reasons discussed in Section~\ref{sec:code_extraction}).

\subsubsection{RQ3}
The most challenging task, requiring both multi-label classification and code analysis, is presented by RQ3. As anticipated, the performance of all models declines compared to RQ1 and RQ2. The highest results are achieved by the most powerful model, GPT-4o, when using method-level data ($C_p$) and the initial issue or pull request description ($R_p$), though the performance is still low, with an F1-score of 0.5839 and an MCC of 0.5928. With complete code ($C_f$), and thus a higher number of observations, the results further drop to an F1-score of 0.4919 and MCC of 0.4805. 

GPT-4o-mini, LLaMA-405B-Instruct, and LLaMA-70B-Instruct have similar performances, but their rankings differ depending on whether the F1-score or MCC is used. For F1-score, from 2nd to 4th place, the order is LLaMA-405B-Instruct, GPT-4o-mini, and LLaMA-70B-Instruct. For MCC, the ranking is GPT-4o-mini, LLaMA-405B-Instruct, and LLaMA-70B-Instruct.
These results suggest that there is considerable room for improvement in this area.

\subsection{Context}
Based on the above discussion, we observe that, in general, $R_f$ aids models in making better decisions for RQ1 and RQ2, but not for RQ3. This is somewhat counterintuitive and may be related to the complexity of RQ3, warranting further investigation. For RQ1 and RQ2, the performance drop is relatively small, indicating that models can still provide practical value when an issue or pull request is initially opened.

Method-level code ($C_p$) appears to yield better results than full code listings ($C_f$), but further analysis is necessary, as the number of observations differs between the two setups.

\section{Threats to Validity}

Validity threats are classified according to~\cite{wohlin2012experimentation,yin2009case}.

\textbf{Internal and construct validity.} Data collection and labeling are error-prone processes. In our study, the potential flaky tests are collected based on cosine similarity and manual labeling. A flaky test can be mislabeled. Our remedy is to have at least two authors cross-examine the potential flaky tests and confirm all positive cases. Non-flaky tests are also collected based on cosine similarity (when its value is less than 0.5) and from those that two authors labeled as non-flaky tests. To mitigate potential errors, at least two authors cross-examined all the non-flaky tests. 

\textbf{External and conclusion validity.}
Generally, software engineering studies suffer from real-world variability, and the generalization problem can only be solved partially~\cite{wieringa2015six}. One threat to external validity is the limited scope of our dataset, which, although enriched from previous studies, still focuses on a subset of quantum software repositories. As a result, the findings may not be fully representative of the broader population of quantum software projects, especially those utilizing different testing frameworks or methodologies. Through future research, we hope to expand our dataset and findings.

\section{Future Plans} 
Our future plans include improving detection methods by exploring and fine-tuning various LLMs, developing automatic patching techniques (given that our dataset includes corresponding fixes), and further exploring the use of LLMs in quantum software maintenance tasks.

\section{Conclusions}
In this paper, we have presented an automated framework for detecting and resolving flaky tests in quantum software, leveraging embedding transformers and LLMs. Our approach enhances the existing dataset of flaky tests and provides methods for semi-automatic and automatic detection. The scores showcase that LLMs perform well, but can be greatly improved for detecting flakiness and root cause detection in quantum software bugs.

\bibliographystyle{IEEEtran}
\bibliography{references}

\end{document}